\documentclass[floatfix, twocolumn,amsmath,amssymb, aps,prb]{revtex4-2}

\usepackage{amsmath}
\usepackage{graphicx}
\usepackage[export]{adjustbox}
\usepackage{caption}
\usepackage{subcaption}
\usepackage{hyperref}
\usepackage{stfloats}

\begin{document}

\title{Transport properties of the parent LaNiO$_2$}
	
\author{Mijanur Islam$^1$, Sudipta Koley$^2$, Saurabh Basu$^1$}

\affiliation{$^1$Department of Physics, Indian Institute of Technology-Guwahati, Guwahati-781039, India.}
\affiliation{$^2$Department of Physics, Amity Institute of Applied Sciences, Amity University Kolkata, Kolkata-700135, India.}

\begin{abstract}
Here we study the transport properties of a nickelate compound, namely LaNiO$_2$ using density functional theory (DFT) in conjunction with dynamical mean field theory (DMFT). An interacting multi-orbital spin-resolved scenario for LaNiO$_2$ yields a metallic ground state with ferromagnetic correlations. The latter contrasts an antiferromagnetic order reported earlier. The metallic behaviour persists even at large values of the interaction energies. Further support of the metallic state is provided by the angular resolved photoemission spectroscopy (ARPES) data.
\end{abstract}
\maketitle

\section{Introduction}
Transport properties of strongly correlated compounds may yield many surprising features, such as non-Fermi liquid behaviour, strong magnetic correlations, the emergence of superconductivity, etc. The nickelates, namely, RNiO$_2$ (R = La, Nd) owing to their prospects of being compared (and contrasted) with the cuprates have started receiving significant attention in recent times. The realization of a superconducting phase in Sr doped NdNiO$_2$ with a transition temperature in the range 9 K - 15 K \cite{Li} rendered support to closer scrutiny on the similarity with the cuprates.\par
The rationale behind drawing a parallel between the nickelates and the cuprates is in the fact that Ni shares a close proximity of Cu in the periodic table. In fact, the infinite layer RNiO$_2$ is isostructural to CaCuO$_2$ \cite{Siegrist}, where the latter represents the parent compound of a high -T$_c$ superconductor. However, the parent compounds of nickelates are reported to be metallic with no magnetic order \cite{Ikeda, Norman, Li}. The absence of magnetic order is attributed to the 3$d$ bands of Ni getting self-populated by holes from the 5$d$ electrons of R resulting in no magnetic order. Indeed, in this sense, the nickelates are quite different than the cuprates. However, the hybridization between the $d_{x^2-y^2}$ orbitals of Ni and those of the rare earths (R) could be quite weak and the self-doping effects may not be dominant.\par
Valence counting of Ni$^{1+}$ cations with a 3$d^9$ configuration is formally similar to Cu$^{2+}$ in the cuprates. The band structure predominantly consists of 3$d_{x^2-y^2}$ orbitals of Ni (similar to Cu $d$-orbitals) and 5$d_{3z^2-r^2}$ orbitals from La with the contribution coming from the oxygen $p$-orbitals to be significantly less than that in the cuprates. In the scenario presented by this, the hopping within the $d$-orbitals of Ni is negligible, and so are the inter-cell hopping amplitudes severely suppressed. In view of this we have only retained the inter-orbital Coulomb repulsion between the 3$d$ bands of Ni with those of 5$d$ of La.\par
Before one understands the effects of doping, a few things need to be settled for the parent compound itself. With a view to accomplish that, we investigate the band structure and the transport properties of LaNiO$_2$ via DFT $plus$ DMFT technique. While we agree on a metallic ground state and absence of any insulating behaviour (unlike cuprates) even at significantly large electronic correlations, our results on the magnetic behaviour disagrees with those in Ref. \cite{Antia}. We distinctly get ferromagnetic correlations due to the Ni atoms via magnetization studies where the magnetic moments of the order of $\sim$ 0.28 $\mu_B$ in the limit of vanishing field. This result severely contradicts the corresponding scenario in cuprates which has an antiferromagnetic insulating ground state for the undoped compound. This certainly holds a promise that the dopant induced superconducting state may yield new physics embedded therein. Further, a test of the familiar Fermi liquid (FL) theory in the context of the metallic ground state is probed by the peak in the single-particle Green's function that can be verified by angular resolved photoemission spectroscopy (ARPES) experiments.\par 
Our paper is organized as follows. We present the DFT band structure along with the crystal structure of nickelate LaNiO$_2$. The density of states (DOS) at the Fermi level is presented to confirm the existence of the metallic behaviour which are further supported by a $T^2$ resistivity that are demonstrated by the experiments \cite{Ikeda, Kaneko}. As a test of the ferromagnetic correlations we present the magnetization data as a function of the applied magnetic field and finally the presence of the quasiparticle peaks provide support to the presence of the FL state.

\section{COMPUTATIONAL DETAILS}
\begin{figure}[htb!]
\centering
\begin{subfigure}[t]{0.45\textwidth}
\includegraphics[width=\textwidth]{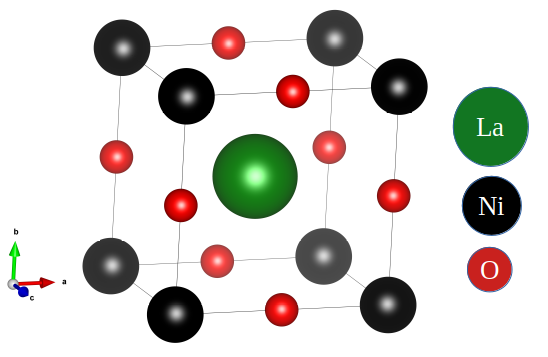}
\caption{}
\end{subfigure}
\begin{subfigure}[t]{0.45\textwidth}
\includegraphics[width=\textwidth]{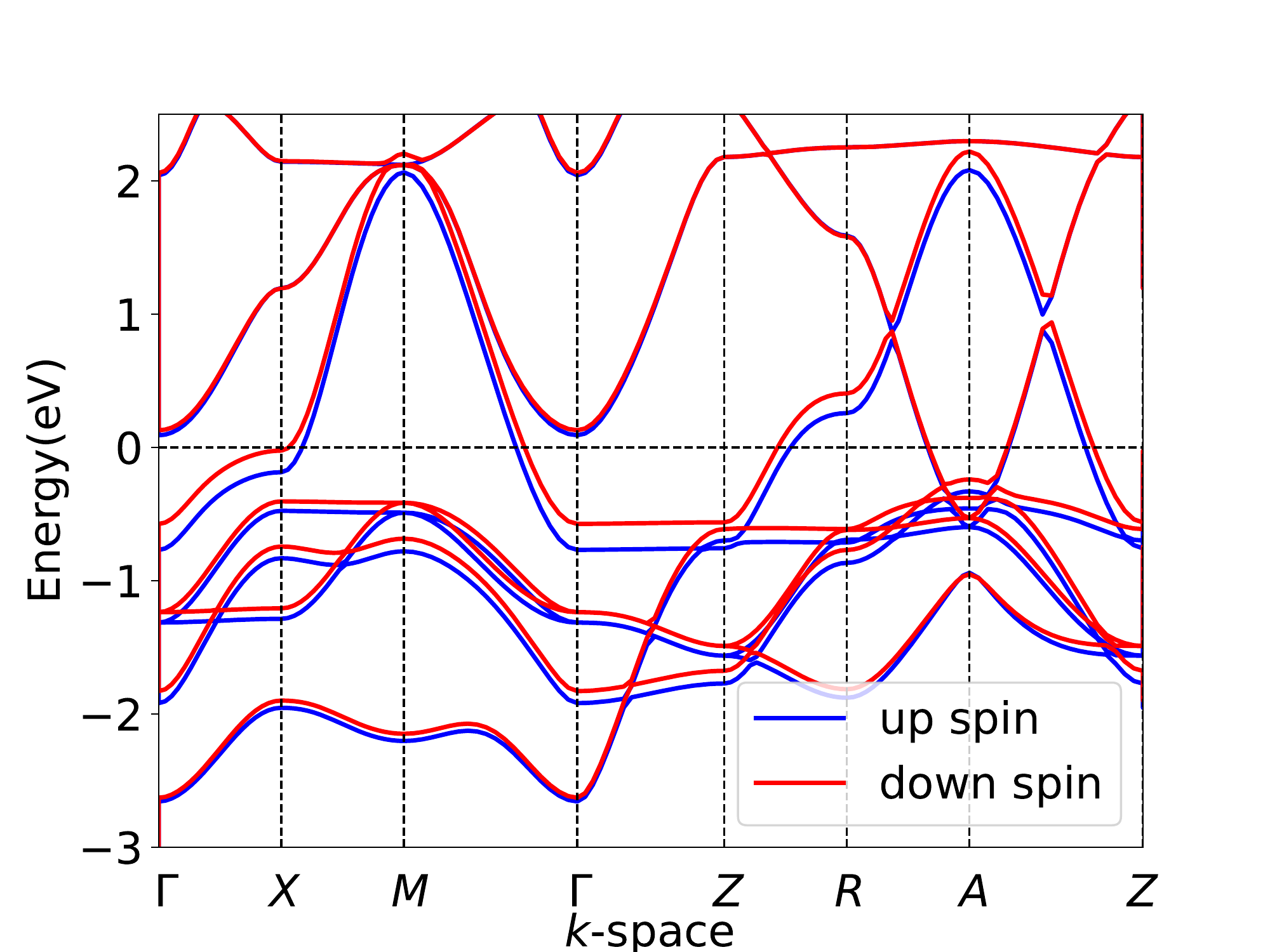}
\caption{}
\end{subfigure}
\caption{(Color Online) (a) Crystal structure of LaNiO$_2$. Green, black and red spheres represent La, Ni and O at- oms respectively. (b) The calculated band structure is shown by DFT within the GGA approximation. The bl- ue and the red lines represent the $\uparrow$- and the $\downarrow$-spin ban- ds respectively.}
\end{figure}

In this work, we have carried out density functional theory (DFT) calculations within Perdew-Burke-Ernzherof (PBE) parameterization of the generalized gradient approximation (GGA) \cite{Perdew} as implemented in Vienna ab initio Simulation Package (VASP) \cite{Kresse} to obtain the band structure and the density of states (DOS) of LaNiO$_2$. Further, we use first-principles DFT combined with dynamical mean field theory to obtain the correlation effects of LaNiO$_2$. For the calculation, the lattice parameters used for LaNiO$_2$ were $a=b=$ 4.002 $\AA$, $c=$ 3.339 $\AA$, and the crystal structure was of the space group 123 ($P4/mmm$). The parameter, R$_{mt} k_{max}$ (R$_{mt}k_{max}$ stands for the product of the smallest atomic sphere radius R$_{mt}$ times the largest $k$-vector $k_{max}$) is chosen to be 7.5, and 10 $\times$ 10 $\times$ 10 $k$-mesh is employed for the structural optimization. Furthermore, the kinetic energy cutoff is set to 500 eV. Finally, the self consistent field (scf) calculations are performed till an energy accuracy of 10$^{-4}$ $eV$ is achieved.\par

 With the DOS from VASP as the input, we employ DMFT via the multi-orbital iterative perturbation theory (IPT) package. We assume the total Hamiltonian to be of the form,
\begin{align*}
H = H_0 + H_{int}
\end{align*}
\begin{multline}
 = \sum_{k,a,\sigma}\epsilon_{k,a} c^\dagger_{k,a,\sigma} c_{k,a,\sigma}+U_1\sum_{i,a} n_{ia\uparrow}n_{ia\downarrow}\\+ U_2\sum_{i,a,b,\sigma,\sigma^\prime} n_{ia\sigma}n_{ib\sigma^\prime} - J_H\sum_{i,a,b}\vec{S}_{ia}\cdot \vec{S}_{ib}
\end{multline}
where $\epsilon_{k,a}$ is the band dispersion, $a$ and $\sigma$ denote the band and spin indices. $U_1$ and $U_2$ are the intra- and inter-orbital Coulomb interaction terms between the electrons with opposite spins at the same site and between electrons with same spins in different sites respectively. $J_H$ is the Hund's coupling. Both being $d$-orbitals, we have chosen the same value of $U_1$ corresponding to the La $d$ and Ni $d$ orbitals. Further, we use the Kanamori relation among the different energy scales, namely 
\begin{equation}
U_2 = U_1 - 2J_H.
\end{equation}
We have considered $J_H = 0.7$ eV, a realistic value for manganites and should work for nickelates as well and varied $U_1$ and $U_2$ over a reasonable range. In keeping with the values of the interaction parameters used in the recent studies \cite{Sakakibara, Nomura}, the results corresponding to $U_1 = 4.0$ eV and $U_2 = 2.6$ eV are presented in our paper. These values were used for the resistivity and the magnetization data, while the spin-polarized DOS includes other values as well. We, however have performed calculation with much larger values of $U_1$ and $U_2$, for example $U_1 = 10$ eV and corresponding $U_2$ defined in Eq. (2). However our results are qualitatively unaltered at very large interaction energies. \par
 
Finally the total Hamiltonian is solved within the multi-orbital iterative perturbation theory (MOIPT) $plus$ DMFT \cite{Dasari, Garg, Koley, Laad, Mohanta, Craco}. IPT is already an established technique for the repulsive Hubbard model in the paramagnetic phase. The IPT technique is constructed in such a way that it should successfully reproduce the leading order terms of the self energy in all of the different cases, such as, weak coupling limit, atomic limit, large frequency limit and it must be exact in the low frequency limit. In the IPT approximation, we approximate the self energy by its second order contribution, namely,
\begin{equation}
\Sigma(\omega) \approx Un + \tilde{\Sigma}^{(2)}_0(\omega)
\end{equation}
here $\tilde{\Sigma}^{(2)}_0(\omega)$ is the second order contribution arising due to the perturbation term, $H_{int}$ (see Eq. 1) and $n$ is the particle number. Though it is an approximate method, it gives good qualitative agreement compared to more exact methods to solve an interacting Hamiltonian.\par
The parameters, namely, the Coulomb interaction, $U_1$ and the Hund's coupling, $J_H$ are varied within an experimentally realizable range to obtain the DOS in presence of electron correlation. Also  $\beta \;(=1/k_BT)$ is the inverse temperature. $T$ is varied up to 300 K to get the temperature dependent behaviour of the resistivity. The DFT $plus$ DMFT self energy calculations are converged up to a precision of 10$^{-4}$ in energy units.

 \begin{figure}[htb!]
\centering
\begin{subfigure}[t]{0.45\textwidth}
\includegraphics[width=\textwidth]{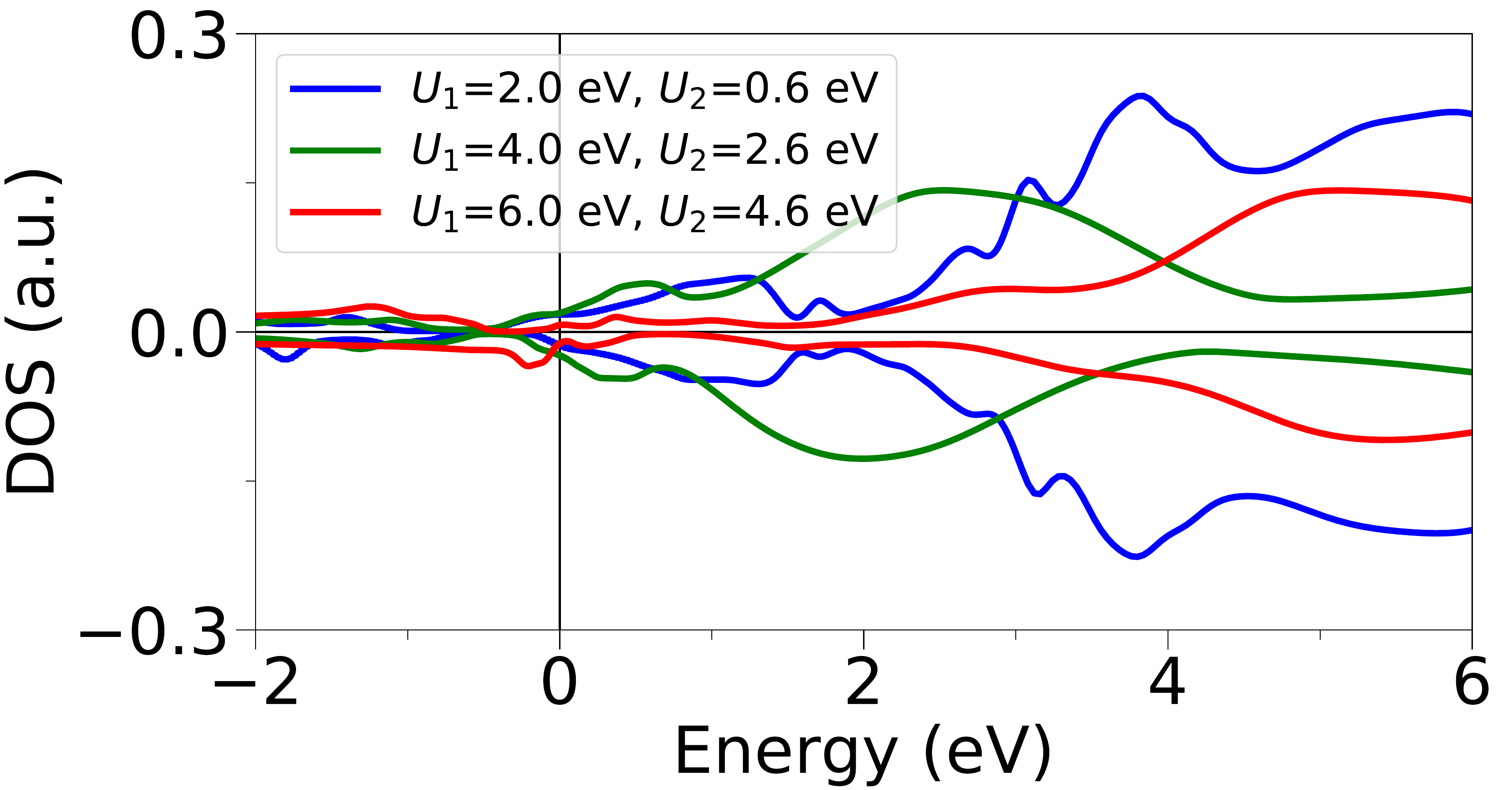}
\caption{}
\end{subfigure}
\begin{subfigure}[t]{0.45\textwidth}
\includegraphics[width=\textwidth]{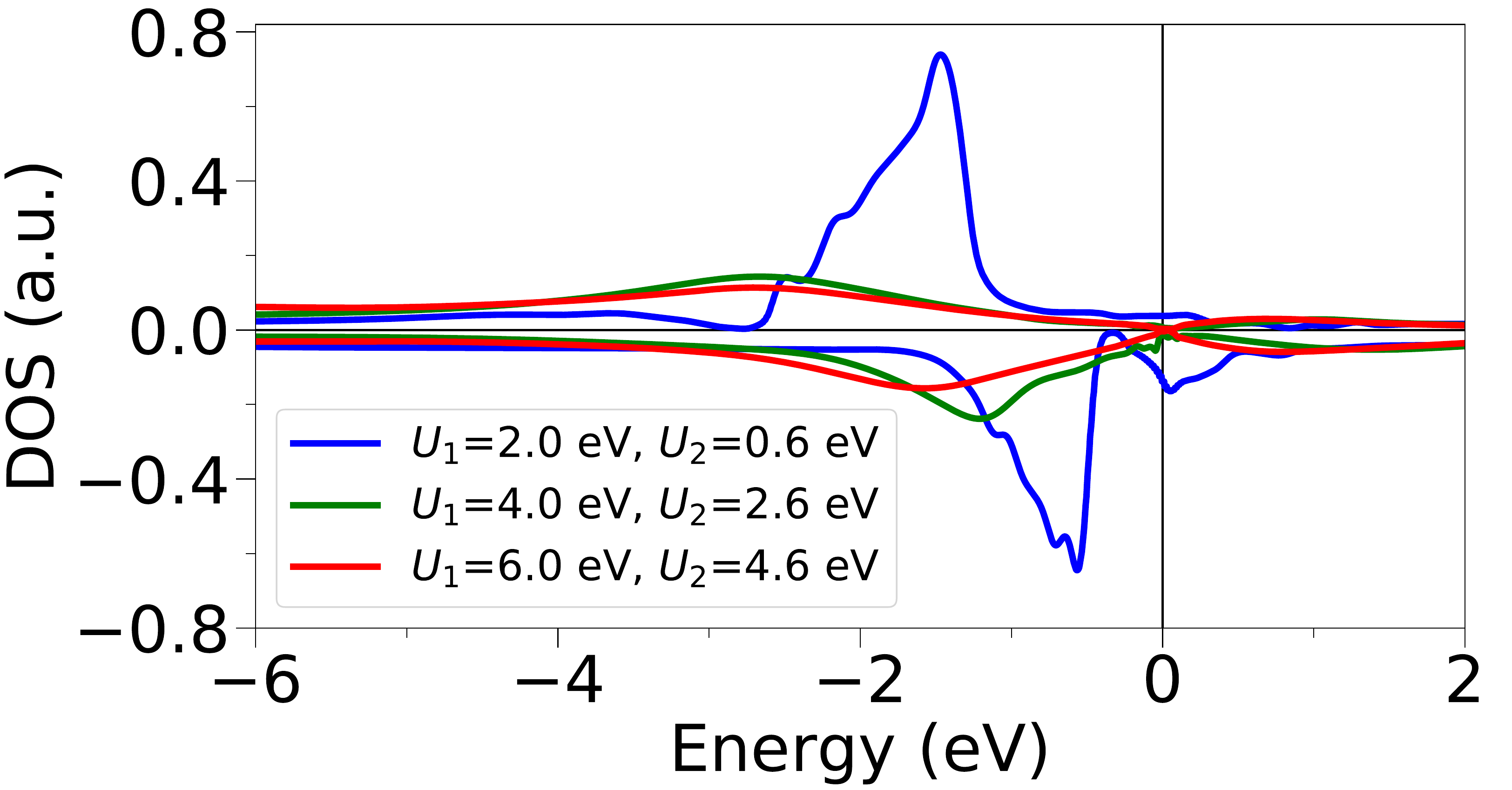}
\caption{}
\end{subfigure}
\begin{center}
\caption{(Color online) The spin-polarized density of st- ates (DOS) for LaNiO$_2$ for various values of $U_1$ and $U_2$ are shown. (a) DMFT DOS for the La atom and (b) DMFT DOS for the Ni atom in LaNiO$_2$. Here positive values of the DOS are presented for the $\uparrow$- spin bands and corresponding to the $\downarrow$- spin bands, the DOS is mu- ltiplied by -1, that is, plotted along the negative $y$-axis for the purpose of presentation.}
\end{center}
\end{figure}

\begin{figure}[htb!]
\centering
\begin{subfigure}[t]{0.45\textwidth}
\includegraphics[width=\textwidth]{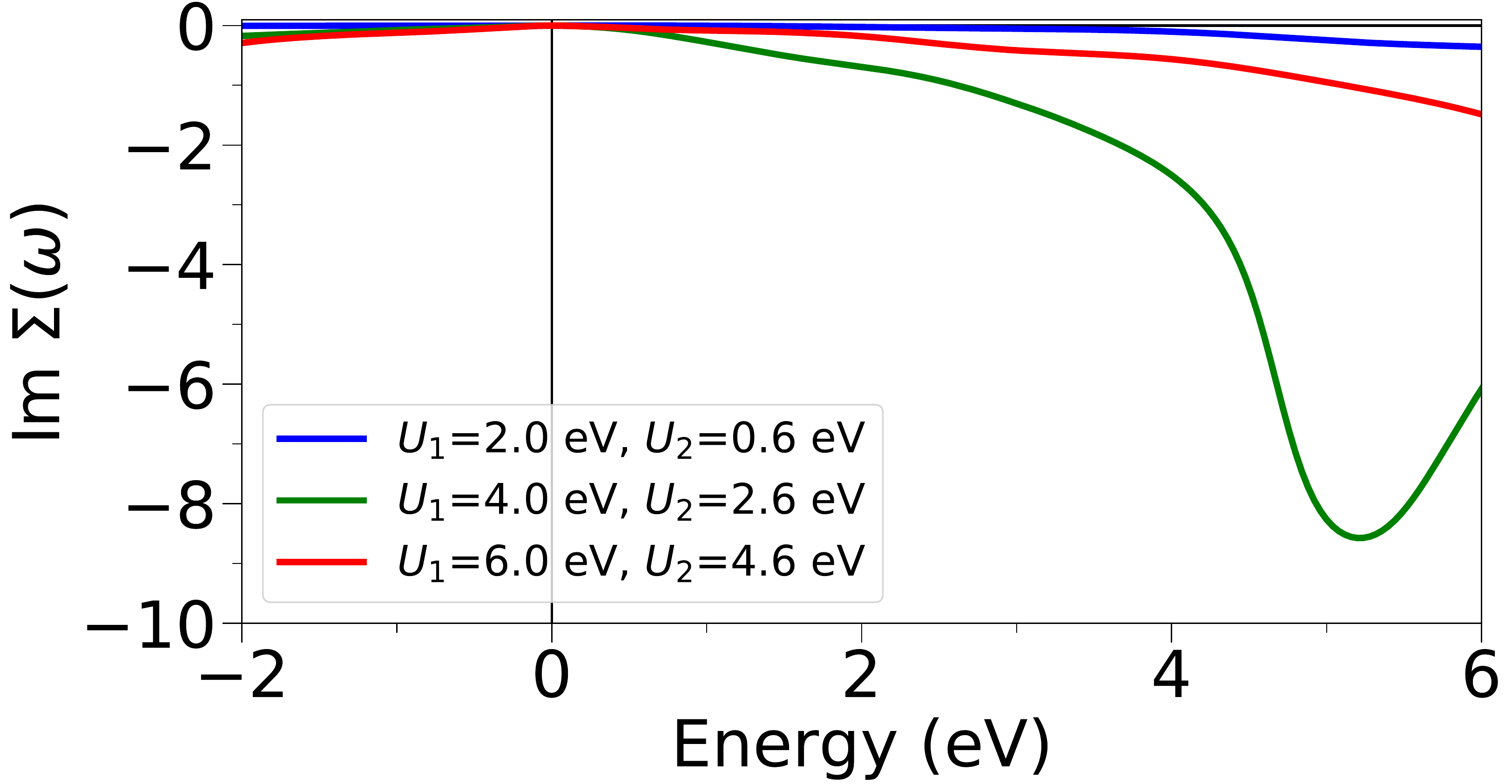}
\caption{}
\end{subfigure}
\begin{subfigure}[t]{0.45\textwidth}
\includegraphics[width=\textwidth]{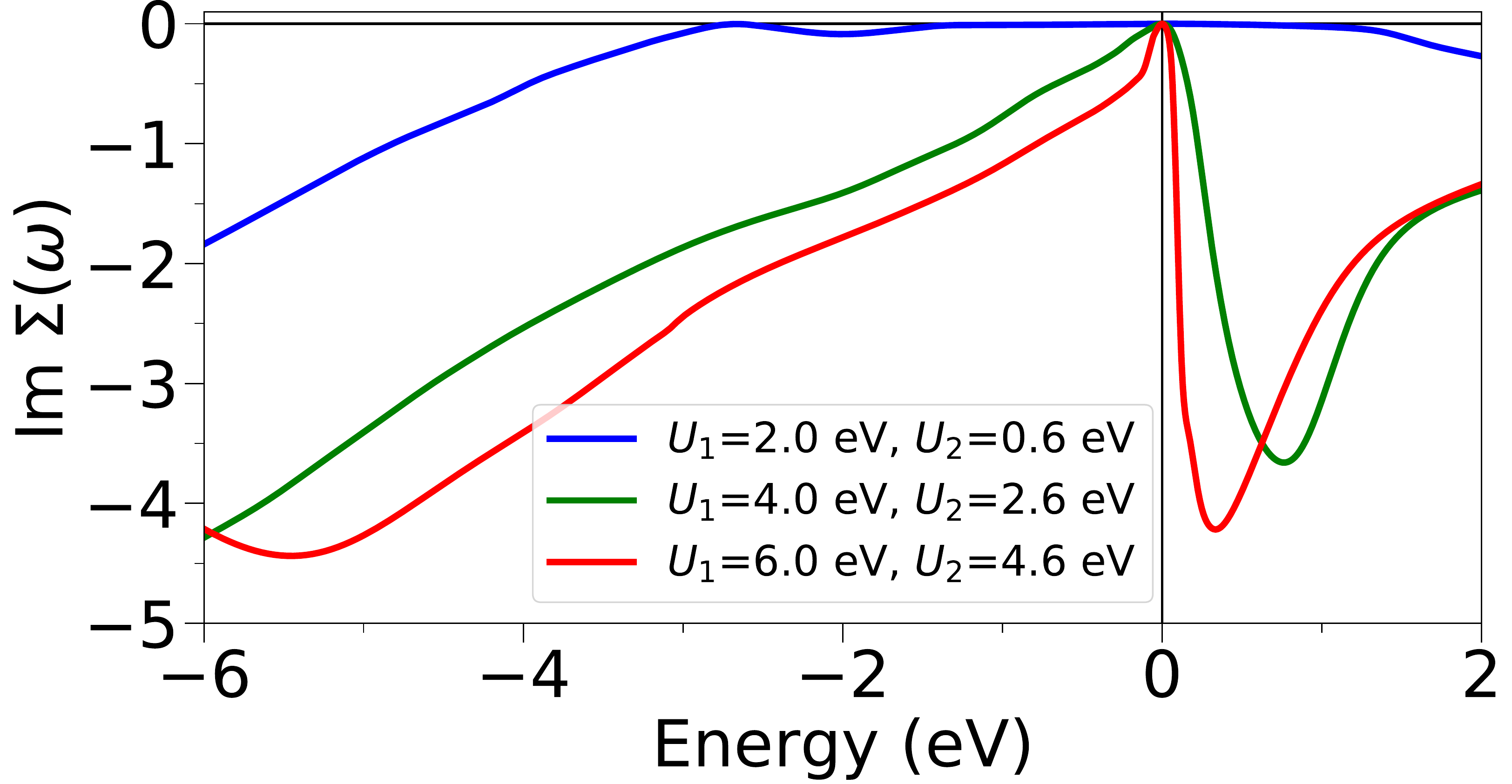}
\caption{}
\end{subfigure}
\begin{center}
\caption{(Color online) Imaginary part of the self-energy (Im$\Sigma(\omega)$) at the same interaction energies as shown in DOS plot. (a) Im$\Sigma(\omega)$ for La $\uparrow$-spin, (b) Im$\Sigma(\omega)$ for Ni $\uparrow$-spin. Here the figures are plotted in the same energy scale as they have been plotted in the DOS figure. We do not show the corresponding data for $\downarrow$-spins as they do not yield any additional information.}
\end{center}
\end{figure}

\begin{figure}
\includegraphics[width=0.5\textwidth]{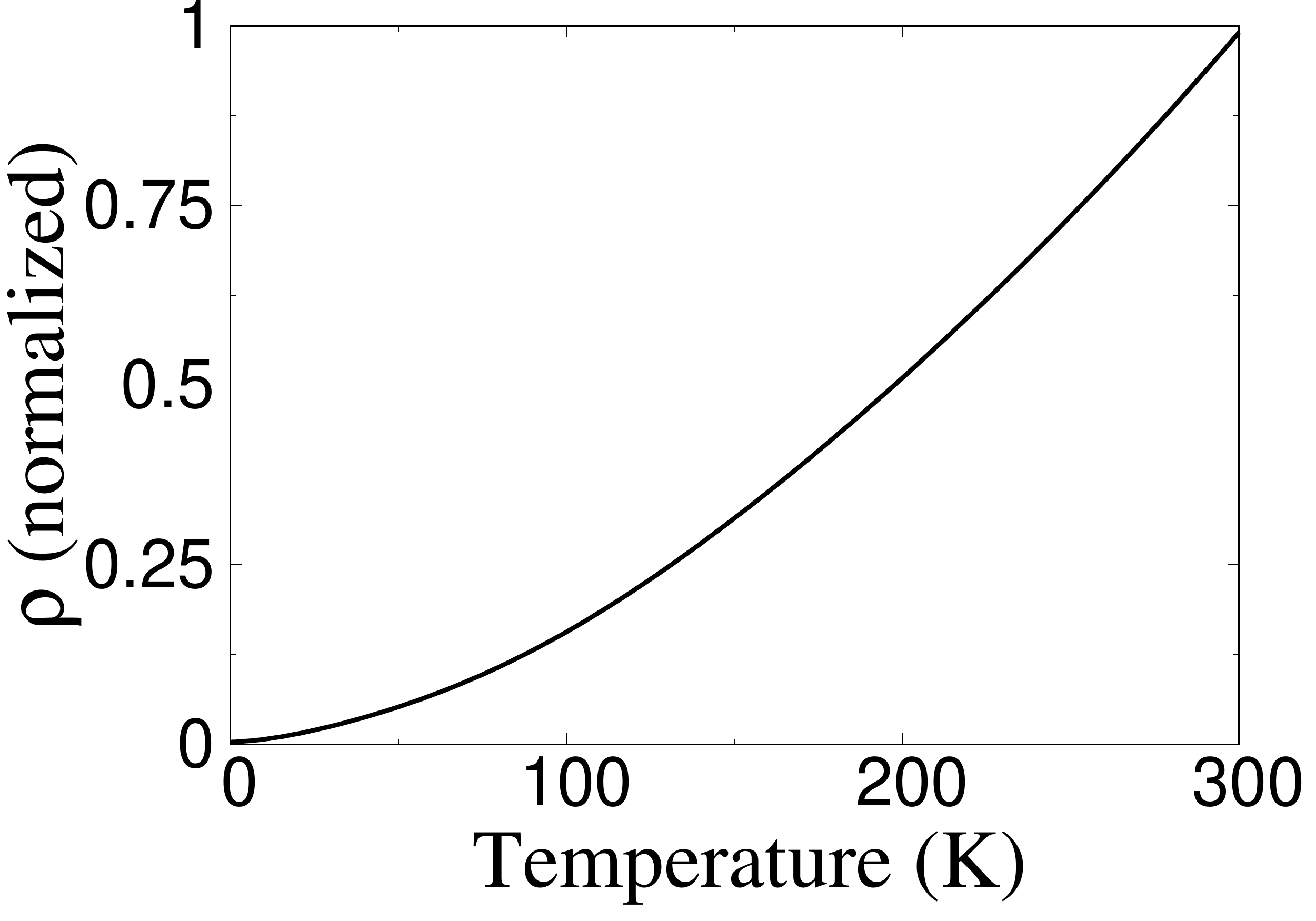}
\caption{(Color online) The DMFT resistivity, $\rho$ (norma- lized) vs temperature ($T$) plot of LaNiO$_2$ at $U_1 = 4.0$ eV and $U_2$ $= 2.6$ eV. Since the $\uparrow$-spin band and the $\downarrow$- spin band show similar behaviour with temperature, na- mely $T^2$, to avoid repetition here we only represent the $\uparrow$-spin resistivity.}
\end{figure}

\section{Results and Discussion}

 In Fig. 1(a), we present the crystal structure of LaNiO$_2$. The crystal structure of LaNiO$_2$ is identical with NdNiO$_2$ (La replacing Nd). Fig. 1(b) represents the spin-polarized DFT band structure of the given material. The blue and the red lines denote the $\uparrow$- and the $\downarrow$-spin bands respectively. We see from our DFT calculation that near the Fermi energy only Ni 3$d$ (3$d_{x^2-y^2}$) and La 5$d$ (5$d_{3z^2-r^2}$) bands are present. The $2p$ bands of oxygen are well separated from them, which is in high contrast with its cuprate counterpart CaCuO$_2$ \cite{Antia}. Now, from the non-interacting density of states obtained from the DFT calculations, we concentrate only on the four bands among all the atomic bands of LaNiO$_2$ that cross the Fermi level. These four bands are La (5$d\uparrow$), La (5$d\downarrow$), Ni (3$d\uparrow$), and Ni (3$d\downarrow$) bands.\par
  The DOS is given by
\begin{equation}
D(\omega)=-\frac{1}{\pi} \rm{Im}G(\omega).
\end{equation}  
 In Fig. 2, we show the orbital-resolved spin-polarized density of states for the La 5$d$ (Fig. 2(a)) and Ni 3$d$ (Fig. 2(b)) atoms computed for various intra- and inter-band interaction energies $U_1$ and $U_2$. The results of the LDA+$U$ calculations on nickelates reported by Anisimov $et$ $al.$ \cite{VI Anisimov} indicated it to be a stable antiferromagnetic insulator. However, the insulating nature of the ground state was contrasted by other groups \cite{Antia, Lee} we too get a metallic behaviour. Though strong electronic correlations renormalize the DOS, the effects are not well visible at the scale presented here. From Fig. 2, it is observed that the spectral weight at the Fermi level, $E_F$ reduces with increasing $U$'s for all the four bands under consideration.\par

 The correlated electronic spectra from the DMFT demonstrates that LaNiO$_2$ is a strongly correlated metal. The metallic behaviour for both the La and the Ni atoms with $\uparrow$- and $\downarrow$-spins were confirmed by vanishing of the imaginary part of the self-energy ($\rm{Im}\Sigma(\omega)$) in the vicinity of the Fermi level ($E_F$). The imaginary part being zero implies the absence of a gap opening at the Fermi level, which, in our case remains robust up to moderate to high values of the interaction potential ($U_1$ and $U_2$). To avoid repetition, we only plot $\rm{Im}\Sigma(\omega)$ corresponding to $\uparrow$-spins for both La and Ni in Fig. 3. The data for $\downarrow$-spins do not yield anything qualitatively distinct. We have checked that the spectral weight continues to be zero at $E_F$. This confirms the correlated metallic behaviour of LaNiO$_2$, which is in good agreement with the previously reported results for nickelates \cite{Lee, Ikeda, Manabe, Antia}. Even at much larger values of $U_1$ (and $U_2$), the system retains its metallic character. Even till as high as $U_1 = 10$ eV, there is no trace of any insulating behaviour that emerges from our calculation.\par

 To further support of the metallic state, we plot the temperature dependence of the dc resistivity defined by $\rho = \frac{1}{\sigma(\omega=0)}$ where $\sigma(\omega)$ is the optical conductivity, which has the form of

\begin{multline}
\sigma(\omega) \propto \frac{1}{\omega} \int_{-\infty}^{\infty} d\epsilon\rho_{0}(\epsilon)\int_{-\infty}^{\infty} d\omega^\prime D_\epsilon(\omega^\prime)D_\epsilon(\omega^\prime + \omega) \\
[n_f(\omega^\prime) - n_f(\omega^\prime + \omega)]
\end{multline}
$\rho_0(\epsilon)$ is the non-interacting DOS and the spectral function is obtained via Eq. 4. For the strongly correlated materials dc resistivity can be calculated using the relation of $\rho$. The parameters used in our calculation are $U_1 = 4.0$ eV and $U_2 = 2.6$ eV and we have the resistivity behaviour from a very low temperature all the way up to the room temperature. We have found that the total resistivity arising out of the contributions from both the bands ($\uparrow$- and $\downarrow$-spin bands) follow a $T^2$ behaviour. Since both the bands overlap with each other, we have only represented the resistivity data for $\uparrow$-spin only. Here the resistivity is normalized with respect to its maximum value. This type of metallic nature of LaNiO$_2$ is consistent with the previously reported data on LaNiO$_2$ \cite{Ikeda, Kaneko}. It is further worthwhile to mention that the dc resistivity data presented in Fig. 4, although comprises of the total contribution from both the La and the Ni bands, the contribution from Ni, that is from the 3$d$ band of Ni is the most dominant one.\par

\begin{figure}[b!]
\centering
\includegraphics[width=0.45\textwidth]{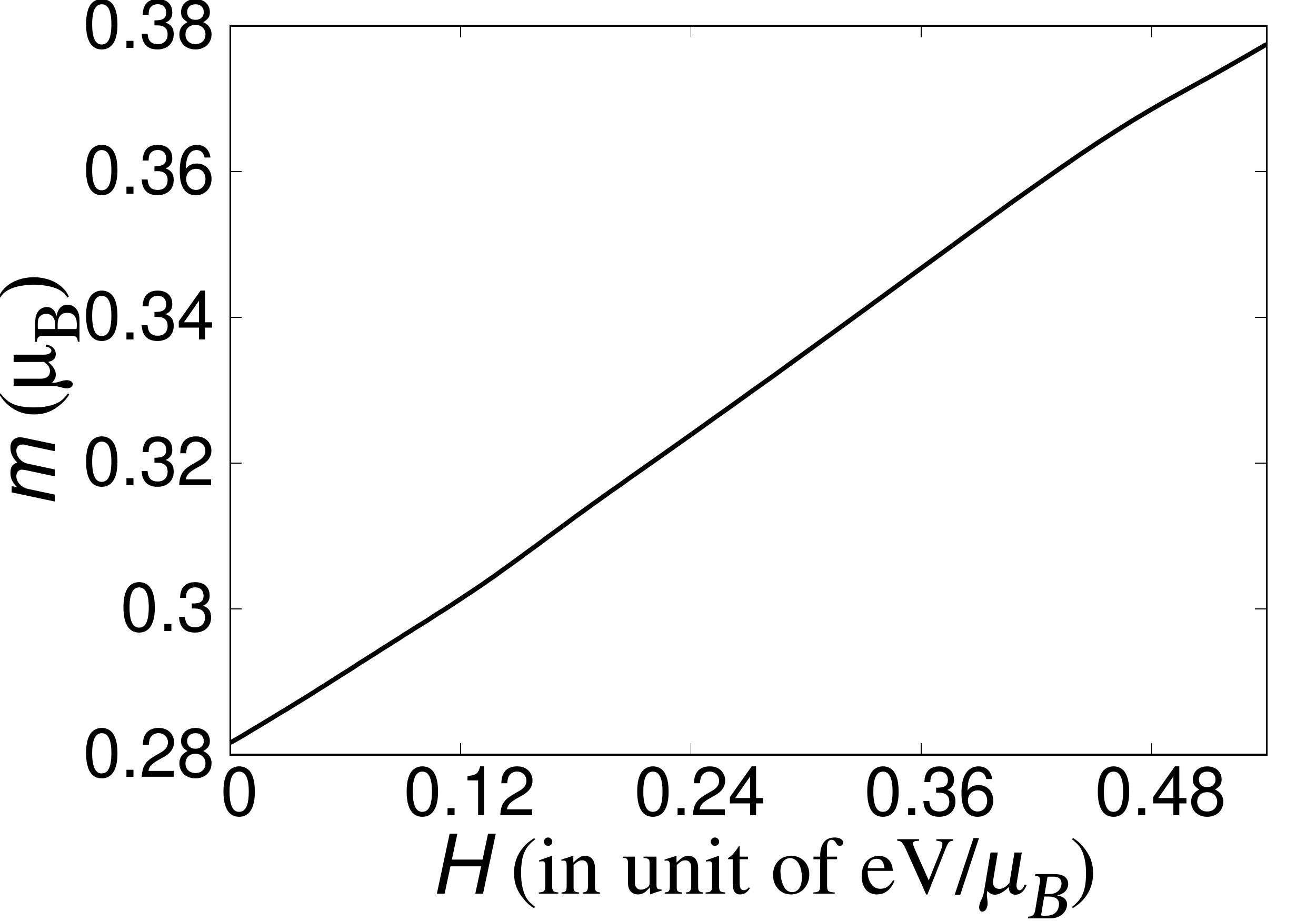}
\caption{(Color online) The magnetization, $m$ as a func- tion of applied magnetic field, $H$ for LaNiO$_2$ at $U_1 =$ $4.$ 0 eV and $U_2 = 2.6$ eV is shown. The value of the magn- etic moment for $H\to 0$ is $\sim 0.28 \mu_B$.}
\end{figure}
 
  In a recent study of the orbital-resolved spin susceptibility for NdNiO$_2$ \cite{Subhadeep}, the importance of the inter-orbital contribution was highlighted. The authors have found that the relative contributions to the susceptibility from the axial $s$-orbital of Ni to that from its $d$-orbital to be only 10$\%$, while the same for the inter-orbital ($s-d$) contribution to the $d$-orbital is 20$\%$. In comparison, the same for LaNiO$_2$ only 1$\%$ and 5$\%$ respectively. This confirms the total susceptibility in LaNiO$_2$ is almost dominated by the $d$-orbital contribution \cite{Subhadeep}.\par

    \begin{figure}[htb!]
\centering
\includegraphics[width=0.45\textwidth]{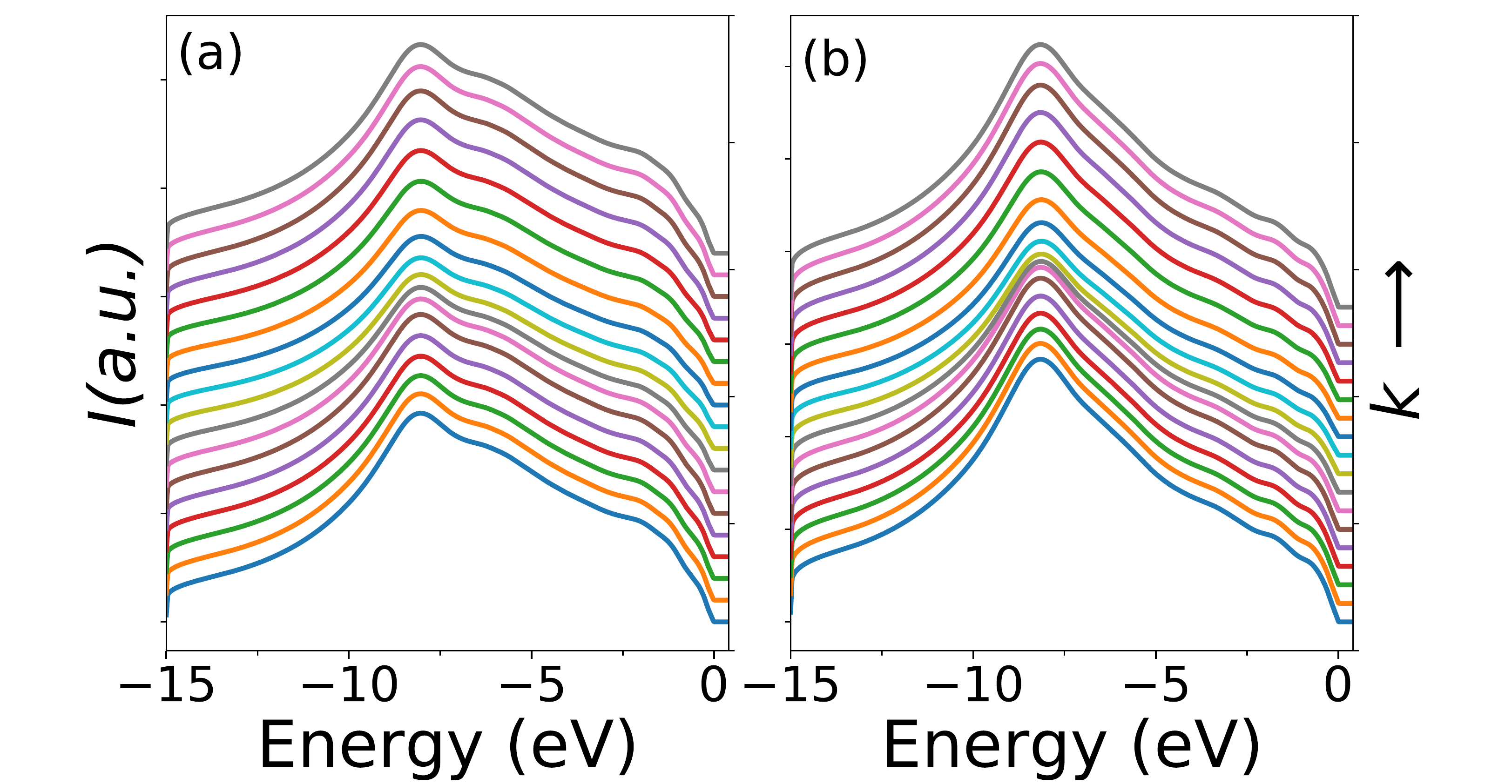}
\caption{(Color online) ARPES energy distribution curv- es (EDC) of parent LaNiO$_2$ along the $\Gamma - X - M - \Gamma$ direction at $T$ = 50 K. (a) for Ni $\uparrow$-spin and (b) for Ni $\downarrow$-spin band. The direction along which the $k$-points in- crease is shown by an arrow.}
\end{figure} 
  
  Next we present the magnetization plot of LaNiO$_2$, where the spin-$\uparrow$ and the spin-$\downarrow$ states are partially filled with an occupation difference between them, thereby resulting in a net magnetic moment. Again the magnetization is predominantly contributed by the Ni 3$d$ orbitals. The magnetization is defined as

\begin{equation}
m = n_{i\uparrow} - n_{i\downarrow}
\end{equation} 
 where $n_{i\sigma}$ are calculated by integrating the spin-resolved DOS. The magnetic moment of Ni is about 0.2831 $\mu_B$, and La atom has a much smaller value, namely 0.00146 $\mu_B$ and the latter also aligns anti-parallel with the moments from the Ni atoms. We have considered the total magnetization that is the magnetization from both the Ni and the La bands.
 
  The system turns out to be ferromagnetic with a value for the magnetic moment to be 0.282 $\mu_B$.  Fig. 5 shows the magnetization, $m$ vs the applied magnetic field, $H$ curve. From the plot, it is clear that $m$ shows a linear behaviour with $H$ in the range shown in Fig. 5. This result is in contrast with the previously reported result of Ref. \cite{Antia}. Where for RNiO$_2$ (R = Nd, La) they have found a stable C-type antiferromagnetic (AFM) ground state with magnetic moments inside the Ni spheres having an approximate value of 0.7 $\mu_B$. However, they have reported a ferromagnetic state which gives rise to a reduced magnetic moment of $\sim$ 0.2 $\mu_B$ at the GGA level, which is less stable than the C-type AFM state by 0.72 $meV$/Ni. The reason behind the discrepancy arises because they have considered a non-interacting ground state, whereas we have considered strong electronic correlations. Also the linear dependence of the magnetization as a function of the external field for undoped LaNiO$_2$ shows widely different behaviour from the corresponding data for cuprates which show insulating long-range antiferromagnetic order in their undoped ground state.\par

 Finally the angle-resolved photoemission spectroscopy (ARPES) spectra can be considered as a helpful technique to support our results for the metallic ground state of LaNiO$_2$. Detailed ARPES studies on LaNiO$_2$ are not available in literature. The ARPES spectral intensity $I(k,\; \omega)$ is given by, $I(k,\omega)=I^0(k)A(k,\; \omega)f(\omega)$ where $A(k,\; \omega)$ represents the spectral function, $I^0(k)$ incorporates the dipole matrix elements, and $f(\omega)$ is the Fermi distribution function. In the case corresponding to a momentum independent self-energy, the spectral intensity takes a form, \cite{LaShell}

\begin{equation}
I(k,\; \omega) \propto \frac{\rm{Im}\Sigma(\omega)}{[\omega - \epsilon_k - \rm{Re}\Sigma(\omega)]^2 + [\rm{Im}\Sigma(\omega)]^2}f(\omega),
\end{equation}
where $\epsilon_k$ is the noninteracting dispersion. $\rm{Im}\Sigma(\omega)$ and $\rm{Re}\Sigma(\omega)$ represent the imaginary and real parts of the self energy respectively. Fig. 6 illustrates the photoemission intensity energy distribution curves (EDC) along $\Gamma - X - M - \Gamma$ direction for the Ni $\uparrow$- and $\downarrow$-spin bands at a temperature $T=50$ K. Fig. 6(a) and Fig. 6(b) show the DMFT single particle spectral function at $T=50$ K for the Ni $\uparrow$- and the Ni $\downarrow$-spin bands respectively. The presence of sharp quasiparticle peaks in the ARPES intensity confirm the Fermi liquid nature of the parent compound LaNiO$_2$.\par

  We would like to make several remarks here. From our DFT $plus$ DMFT calculations, we show that the normal state of the parent LaNiO$_2$ is quite different from its copper oxide counterpart. The ground state of LaNiO$_2$ is in a strongly correlated metallic state, which agreed and also have conflicts with some previously reported data. We have also identified the Ni-3$d$ (3$d_{x^2-y^2}$) orbital as the most correlated one. The metallic nature was confirmed by the $\rm{Im}\Sigma(\omega)$ data and also LaNiO$_2$ shows a $T^2$ behaviour in the resistivity vs temperature plot. Further, LaNiO$_2$ follows a linear dependence of magnetization, that is a ferromagnetic behaviour upon applying an external magnetic field. The magnetic ordering of the parent nickelate LaNiO$_2$ is quite different from the magnetic ordering of cuprates. In addition, in support of our Fermi liquid ground state, we also provide the ARPES data. The sharp quasiparticle peaks in ARPES intensity confirm the presence of a FL ground state.

\end{document}